\magnification=\magstep1
\centerline{\bf On the Minimum Energy Configuration of a Rotating
Barotropic Fluid:}
\centerline{\bf A Response to Narayan \& Pringle astro-ph 0208161}

\bigskip
\bigskip
\centerline{S\'ebastien Fromang (IAP) \& Steven A. Balbus (UVa)}
\bigskip
\bigskip
\noindent

\medskip

\noindent In a recent posting, Narayan \& Pringle (2002,
astro-ph/0208161) criticized a paper written by the current authors
(Fromang \& Balbus 2002,  astro-ph/0207561 hereafter FB02).  FB02
presented a second order variational calculation and concluded that
supersonically rotating barotropic fluids had lower energy states
neighboring that of equilibrium.  (Note that this is not the same as
claiming instability, a point noted explicitly in FB02.)  Narayan \&
Pringle challenged this, presented a specific class of equilibrium
solution, and claimed to prove that it represented a global
energy minimum.  They concluded that FB02 was incorrect.

The calculation presented in FB02 is indeed incorrect.  The difficulty
is that the second order changes in the energy cannot be calculated
directly from the energy itself, even when the variations are
restricted to satisfy the constraints of mass and angular momentum
conservation.  Rather, the constraints themselves change the effective
functional form of the energy, by forcing additional second order
correlations that would otherwise be absent.  Narayan \& Pringle (2002)
is therefore correct in stating that our second order variational
calculation is in error.

Nevertheless, we are compelled to respond to the critique of Narayan \&
Pringle (2002).   Our reasons are twofold: (1) the critique unfairly
misrepresents statements in FB02; and (2) the main discussion presented
in the critique appears itself to have flaws.  In particular, the
conclusion of Narayan \& Pringle that their chosen example is a global
minimum with respect to arbitrary variations from equilibrium does not
follow from the argument presented in their paper.  We very briefly
discuss these two points in turn.

\bigskip

\noindent {\it 1. Lower energy states and instability.\/} The existence
of lower energy states neighboring the equilibrium solution is not a
sufficient criterion for instability.  This point is stated explicitly
in FB02:

\medskip
\noindent ``Note that our procedure does not ensure instability in the
latter case [supersonic rotation], it simply shows that neighboring
states of lower energy exist.'' \medskip

\noindent Unfortunately, Narayan \& Pringle misrepresent FB02,
stating, for example:

\medskip
\noindent ``Since FB02 claim that the only stable barotropic
configurations are those that rotate uniformly {\it and\/} are subsonic
at all radii, $...$'' 

\medskip

\noindent Throughout the critique, the existence of lower energy states
is referred to as a ``stability criterion.''  No such claim was ever
made in FB02.  A stability analysis of a uniformly rotating polytrope
in an external potential need not show any unstable modes, any more
than a stability analysis of, say, a Keplerian disk is bound to show
linear instability, even though such a profile is not an energy
minimum.  When a weak magnetic field is added to a Keplerian disk, no
matter how small the field energy, pathways are opened to states of
lower disk energy and the MRI ensues, even for purely axisymmetric
disturbances.  In the case of a rotating polytrope, the question is not
whether the system itself is directly unstable {\it per se,\/} it is
whether subsequent evolution drives the initial state further away from
its original equilibrium when one makes energetically negligible
modifications --- say a weak magnetic field and a tiny angular velocity
gradient.

\medskip \noindent

{\it 2. Global Energy Minima.}  Narayan \& Pringle analyze rotating
polytropes in the fixed external potential
$$
\Phi = {1\over 2} \Omega_K^2 (R^2 + z^2)
$$
where $\Omega_K$ is a constant and $R$ and $z$ are cylindrical radius
and vertical coordinate respectively.  For a given mass $M$ and angular
momentum $J$, it is shown that there is a unique equilibrium solution.
Combining this result with the positive-definite character of the
energy $E$, Narayan \& Pringle conclude that the energy extremum
represented by their solution must be a global minimum.

Clearly something more is required to establish a global minimum.  This
line of reasoning is insufficient to establish the existence of a {\it
local\/} minimum, let alone a global one.  (A gaussian function is
everywhere positive-definite and possesses a unique extremum, which is,
of course, a maximum.)  We do not wish to belabor this issue or other
technical difficulties with the presentation, since the point of
discrediting the variational calculation is moot.  The point of
physics, however, is not without interest.

Consider the gravitational potential arising from a constant density
sphere of large but finite radius.  Within the sphere, the potential is
exactly of the above form (perhaps with an inconsequential additive
constant), while outside the sphere the potential is Keplerian.  There
will be values of $M$ and $J$ for which the unique equilibrium solution
lies entirely in the sphere, and is of the precise form considered by
Narayan and Pringle.  The total energy is bounded from below (and can
thus be made positive by adding a trivial constant to the potential).
But for any equilibrium solution lying entirely within the sphere,
there exists another state with the same $M$ and $J$ that has a lower
energy: one in which the mass distribution is essentially identical,
and the angular momentum is removed to a ring of arbitrarily small mass
and arbitrarily small total energy at arbitrarily large distances.
(Supersonic disks do in fact avail themselves of these lower energy
solutions.)  Note that if the potential rises sufficiently steeply at
large distances, the argument need not hold.  Thus, the nature of the
potential at large $r$ plays a role in establishing the existence of
global minima.  If the solutions of Narayan \& Pringle (2002) are in
fact true global minima, this is dependent upon the rapid growth of the
harmonic potential at large distances, and is not a simple consequence
of the existence of a unique extremum and positivity of the energy.

\bye